\documentclass[aps,onecolumn,pre,10pt,groupedaddress,notitlepage]{revtex4-1}

\usepackage{graphicx}  
\usepackage{dcolumn}   
\usepackage{bm}        
\usepackage{amssymb,amsfonts,amsmath}   
\usepackage{color}
\usepackage{lipsum,subeqnarray}

\usepackage[normalem]{ulem}

\allowdisplaybreaks

\def\Xint#1{\mathchoice
{\XXint\displaystyle\textstyle{#1}}%
{\XXint\textstyle\scriptstyle{#1}}%
{\XXint\scriptstyle\scriptscriptstyle{#1}}%
{\XXint\scriptscriptstyle\scriptscriptstyle{#1}}%
\!\int}
\def\XXint#1#2#3{{\setbox0=\hbox{$#1{#2#3}{\int}$}
\vcenter{\hbox{$#2#3$}}\kern-.5\wd0}}
\def\ddashint{\Xint=}

\begin{document}

\title{A compact Eulerian representation of axisymmetric inviscid vortex sheet dynamics}

\author{Adriana~I.~Pesci$^{1}$, Raymond~E.~Goldstein$^{1}$, and Michael~J.~Shelley$^{2}$}
\affiliation{$^{1}$Department of Applied Mathematics and Theoretical Physics,\\Centre for Mathematical Sciences, University of 
Cambridge, Wilberforce Road, Cambridge CB3 0WA, United Kingdom\\
$^{2}$Courant Institute of Mathematical Sciences, New York University\\ 251 
Mercer Street, NY 10021, USA}

\date{\today}

\begin{abstract} 
A classical problem in fluid mechanics is the motion of an axisymmetric vortex sheet
evolving under the action of surface tension, surrounded by an inviscid fluid.  Lagrangian
descriptions of these dynamics are well-known, involving complex nonlocal expressions for the
radial and longitudinal velocities in terms of elliptic integrals.  Here we use these prior
results to arrive at a remarkably compact and exact Eulerian evolution equation for the sheet
radius $r(z,t)$ in an explicit flux form associated with the conservation of enclosed volume.
The flux appears as an integral involving the pairwise mutual induction formula for 
vortex loop pairs first derived by Helmholtz and Maxwell.  We show how the well-known linear stability
results for cylindrical vortex sheets in the presence of surface tension and streaming flows 
[A.M. Sterling and C.A. Sleicher, \textit{J. Fluid Mech.} {\bf 68}, 477 (1975)]
can be obtained directly from this formulation.
Furthermore, the inviscid limit of the empirical model of Eggers and Dupont 
[\textit{J. Fluid Mech.} \textbf{262} 205 (1994); \textit{SIAM J. Appl. Math.} {\bf 60}, 1997 (2000)], which has served as the basis
for understanding singularity formation in droplet pinchoff, is derived within the present formalism as the leading 
order term in an asymptotic analysis for long slender axisymmetric vortex sheets, and should provide the
starting point for a rigorous analysis of singularity formation.
\end{abstract}
\maketitle

\section{introduction}
\label{sec:intro}

\begin{figure}[t]
\centering
\includegraphics[width=0.95\textwidth]{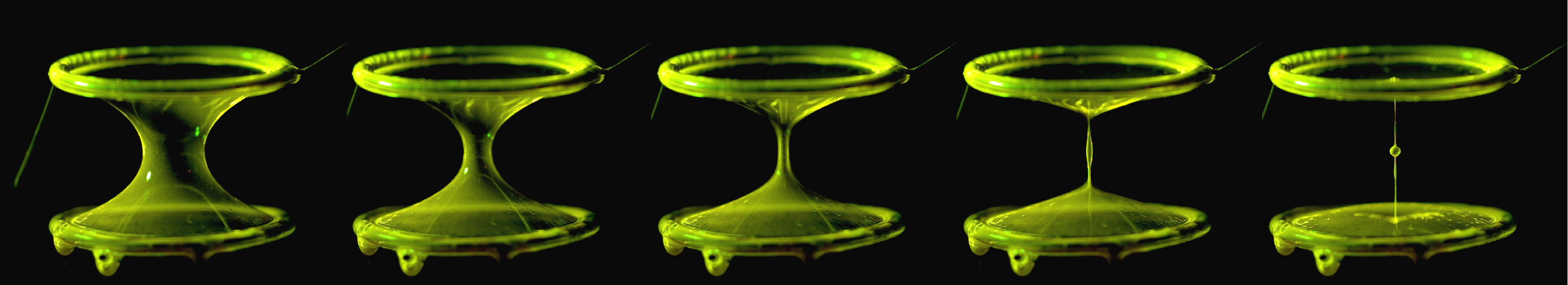}
\caption{Collapse of a catenoid.}
\label{fig1}
\end{figure}

Perhaps the simplest example of finite-time singularities in surface motion is that which occurs when the two
loops supporting a catenoid-shaped soap film are pulled apart beyond a critical separation, rendering the film
unstable.  Extensive experimentation \cite{RobinsonSteen} has shown that the collapsing axisymmetric surface eventually breaks up 
through self-contact at multiple points, producing a series of satellite soap bubbles (Fig. \ref{fig1}). Theoretical work on this
dynamical process ranges from Maxwell's original stability analysis \cite{Maxwell_catenoid} to
much more recent computational studies of fluid dynamical models \cite{ListerHinch,ChenSteen} focused on
the nature of the singularities.

In a broader sense, there have been two schools of thought in the study of singularity formation by moving surfaces.  On the one hand,
there is a very substantial body of rigorous work on the simplest geometrical law, namely 
motion by mean curvature \cite{mean_curvature_flow}.  However, while this law is a physically realistic description 
of interface motion arising from surface diffusion, it cannot be applied to the motion of soap films because
it can not account for the dynamics of the surrounding fluid and the conservation laws that follow.
There are modifications of mean curvature flow that conserve volume enclosed by the surface, but they are not 
faithful representations of the dynamics of the surrounding fluid.
On the other hand, there is the large body of more phenomenological work on singularity formation in fluid mechanics
\cite{eggers_review}, where simplified PDE models have been developed to address
self-similar dynamics near singularities.  These models are physically realistic but have generally
lacked rigor.

The two examples where the gap between these two approaches has been bridged is in the context of interface
motion in two dimensions, where a systematic procedure to derive PDEs for the evolution of asymptotically thin fluid layers
has been developed from the exact boundary-integral formulation 
\cite{GPS_PRL,PughShelley,GPS_PRL2,GPS_PF}.  These analyses have put on a solid foundation empirical models 
\cite{Constantin,Dupont,EggPont}
derived within lubrication theory.
Given these results, the question then arises of whether there is a comparable physical setup in three dimensions in which 
a PDE can be derived for the motion of a surface surrounded by an incompressible fluid. The simplest example of this
would naturally be an axisymmetric surface surrounded by an inviscid fluid.  
If there is an affirmative answer to this question, such a PDE should have an explicit flux form.

In the inviscid limit, the moving interface can be represented by a vortex sheet with surface tension.
The predominant approach to this problem has been through a Lagrangian formulation \cite{Moore,CaflischLi}, which is the most 
appropriate for computational studies \cite{BakerOrzag,NieBaker,Nie,MikeJohn}. However, such an approach does not readily lend
itself to the development of asymptotic models appropriate to thin necks, as would be relevant in the neighborhood
of singularities.  In contrast, an Eulerian formulation would not only be amenable to asymptotic analysis, but would also 
be subject more easily to rigorous studies.

Here, as a first step, and starting from the Lagrangian formulation, we derive an exact Eulerian dynamics for 
axisymmetric vortex sheets with surface tension, and show that it has an explicit flux form.  Naturally, 
because the problem is deeply nonlocal due to the Biot-Savart interactions between distant elements of the vortex sheet,
the flux is an integral over the entire sheet whose kernel is precisely the mutual induction between two coaxial loops as
derived by Helmholtz for fluids \cite{Helmholtz_loops} and by Maxwell for electrical currents \cite{Maxwell_loops}.  As
the mutual induction is, by definition, the flux through one loop due to the circulation in the other, this is a 
very intuitive result.  We show that known stability results for cylindrical vortex sheets can be recovered by direct
calculation.
While nonlocal, the mutual induction between two coaxial loops of the same radius 
is sharply peaked as the distance between the loops vanishes, and this feature suggests a natural asymptotic analysis to 
reduce systematically the dynamics to a local PDE.  In this limit we recover at first order 
the inviscid version \cite{EggSIAM} of the empirical model first proposed by Eggers and Dupont \cite{EggPont}
for this problem.  

\section{Flux Equation}
\label{sec:flux}

We consider an infinite three-dimensional inviscid fluid of density $\rho$ within which is a vortex sheet that is axisymmetric
about the $z$ axis, and whose time-dependent radius is $r(z,t)$.    In this
inviscid limit there is in general a discontinuity in the tangential fluid velocity across the sheet, and this
jump defines the vortex sheet strength $\gamma$.  
In the same way that we use a minimal surface as an idealization of a static soap film, here the vortex sheet is
the idealized representation of a moving film endowed with 
surface tension $\sigma$.  As is well-known \cite{CaflischLi}, the problem of the self-induced motion of the sheet 
reduces to the coupled dynamics of $r(z,t)$ and $\gamma(z,t)$, for which the vortex sheet strength evolves as
\begin{equation}\label{strength}
{\partial \gamma \over \partial t} = -{\sigma \over \rho} {\partial \kappa \over \partial \alpha},
\end{equation}
where $\kappa $ is the mean curvature of
the surface, 
\begin{equation}
\kappa=\frac{r_{zz}}{\left(1+r_z^2\right)^{3/2}}-\frac{1}{r\left(1+r_z^2\right)^{1/2}}~,
\end{equation}
and
the radial and axial Lagrangian
velocities are the principal value integrals
\begin{subequations}
 \begin{align}
z_t\equiv {dz\over dt} &= -{1 \over 4 \pi}  \ddashint \gamma^{\prime} \left[ {r^{\prime}}^2  J_2 - r r^{\prime} J_1 \right] d\alpha^{\prime},
\label{CL1a} \\
r_t \equiv {dr\over dt} &= -{1 \over 4 \pi}  \ddashint \gamma^{\prime} r^{\prime} (z - z^{\prime}) J_1 d\alpha^{\prime},
\label{CL1b}
\end{align}
\label{CL1}
\end{subequations}
where 
$\alpha'$ is a Lagrangian label, $r'\equiv r(\alpha',t)$, $\gamma'\equiv \gamma(\alpha',t)$, and
\begin{subequations}
\begin{align}
J_1 &=  \int_0^{2\pi} \cos \theta^{\prime} \left[ (z - z^{\prime})^2 + r^2 + 
{r^{\prime}}^2  - 2 r r^{\prime}\cos \theta^{\prime} \right]^{-3/2} d\theta^{\prime},\label{J1define} \\
J_2 &= \int_0^{2\pi} \left[ (z - z^{\prime})^2 + r^2 + 
{r^{\prime}}^2  - 2 r r^{\prime}\cos \theta^{\prime} \right]^{-3/2} d\theta^{\prime}.
\label{J2define}
\end{align}
\label{Jdefine}
\end{subequations}
Performing the integrations in $\theta'$ yields the following expressions for the velocities in terms of 
$K$ and $E$, the complete elliptic integrals of the first and second kind, respectively, 
with argument $k^2 =  4 r r^{\prime}/\left[(z - z^{\prime})^2 + (r +r^{\prime})^2\right]$,
\begin{subequations}
 \begin{align}
z_t &= {1 \over \pi}  \ddashint d\alpha' \gamma' r' \left( {k^2 \over 4 r r'}\right)^{3/2}\left[ {r' - r \over 1-k^2}
E(k) -{2r\over k^2} \left[ E(k) - K(k) \right] \right] \equiv \ddashint d\alpha' \gamma' P(\alpha,\alpha') \label{CL2a} \\
r_t &= {1 \over \pi}  \ddashint d\alpha' \gamma' r' (z - z') \left( {k^2 \over 4 r r'}\right)^{3/2}\left[ {2\over k^2}
 \left[ E(k) - K(k) \right]  + {1 \over 1-k^2} E(k) \right] \equiv \ddashint d\alpha' \gamma' Q(\alpha,\alpha').
 \label{CL2b}
 \end{align}
 \label{CL2}
\end{subequations}

In order to obtain an Eulerian equation of motion
from these Lagrangian velocities we make the change of variables from $\alpha$ to $z=z(\alpha,t)$.
This yields \cite{PughShelley}
\begin{subequations}
 \begin{align}
  {\partial r \over \partial t} &= r_t- z_t {\partial r \over \partial z}\label{strength1a} \\
  {\partial \gamma \over \partial t}+z_t {\partial \gamma \over \partial z} &= 
  -{\sigma \over \rho}z_{\alpha} {\partial \kappa \over \partial z}.
  \label{strength1b}
 \end{align}
 \label{strength1}
\end{subequations}
The quantity 
$\tilde\gamma=\gamma/s_{\alpha}$ represents the actual jump in tangential velocity \cite{Peng,Herrmann}, where $s$ is the arclength, whereas it is the unnormalized 
vortex sheet strength 
$\gamma$ that appears in  Eqs. \ref{CL2} and \ref{strength1}.  For an Eulerian description 
in which we make the change of variables from $\alpha$ to $z=z(\alpha,t)$, it is 
natural to consider instead the quantity $\eta=\gamma/z_{\alpha}$ , so
$d\alpha' \gamma'=dz'\eta'$.

For this system, conservation of fluid volume should be expressible as a flux form involving the cross-sectional
area $\pi r^2$, that is,
\begin{equation}
{\partial r^2 \over \partial t} = -{\partial F \over \partial z} ,
\label{flux}
\end{equation}
for some function $F$ which we now seek.  The fact that $2r\partial r/\partial t=\partial r^2/\partial t$, implies that
$2r(r_t-z_t\partial r/\partial z)=-\partial F/\partial z$.  However, to infer $F$ by direct substitution of \eqref{CL2a} 
and \eqref{CL2b} into this expression 
is cumbersome and non-trivial.  A much better way to find $F$ is to make use of Eqs. \eqref{CL1} and \eqref{Jdefine}.
After some algebra and integrations by
parts, we find
\begin{equation}\label{flux2}
F =  {1\over 2 \pi }\ddashint \eta^{\prime} r r^{\prime} dz^{\prime}
\int_0^{2\pi} { \cos \theta^{\prime} \over \left[ (z - z^{\prime})^2 + r^2 + 
{r^{\prime}}^2  - 2 r r^{\prime}\cos \theta^{\prime} \right]^{1/2}} d\theta^{\prime}.
\end{equation}
The integral over the variable $\theta^{\prime}$ can be calculated in terms of elliptic integrals, with the result
\begin{equation}\label{fluxMaxwell}
{\partial r^2 \over \partial t} =  -{1\over  \pi } {\partial \over \partial z}\int dz^{\prime} 
\eta^{\prime} {\sqrt {r r^{\prime}} }
M(k)~,
\end{equation}
where $M(k)$ is Maxwell's function first derived for the mutual induction of a pair of circular current loops of radii $r$ and $r'$
at locations $z$ and $z'$,
\begin{equation}\label{Mfunction}
M = {2 \over k} \left(K - E  - {k^2 \over 2} K\right).
\end{equation}
That this remarkably compact result has not been previously obtained in this context may be a consequence of the 
natural emphasis on the Lagrangian formulation and its computational applications.  The results in \eqref{fluxMaxwell} and \eqref{Mfunction}
are intuitive, in that the mutual induction $M$ is, by definition, the flux passing through one loop due to another, and so
$F$ is simply the sum of all the individual fluxes through the loop at position $z$.  Moreover, in the case of a vortex loop, the
integrand in $F$ was identified by Helmoltz \cite{Helmholtz_loops} as the stream function of the flow.

In closing this section, we note that if we rewrite the pair \eqref{CL2} as
$z_t = W$ and $r_t=V$, then $\eta_t+W_z\eta=-(\sigma/\rho) \kappa_z$.  Considering $\eta(\alpha,t)=\tilde\eta(z(\alpha,t),t)$, and similarly
$r(\alpha,t)=R(z(\alpha,t),t)$, the Eulerian evolution equations for $\eta$ and $R$ that parallel \eqref{strength1} are
\begin{subequations}
 \begin{align}
  {\partial R \over \partial t} &= V- W {\partial R \over \partial z}\label{Mikea} \\
  {\partial \tilde\eta \over \partial t}+{\partial \over \partial z}\left(W\tilde\eta\right) &= -{\sigma \over \rho} {\partial \kappa \over \partial z}.
  \label{Mikeb}
 \end{align}
 \label{Mike}
\end{subequations}
Multiplying \eqref{Mikeb} by $R$ and integrating the incompressibility relationship $(1/r)(rV)_r+W_z=0$, one can easily show
\begin{equation}
 \frac{\partial}{\partial t}\left(\frac{1}{2}R^2\right)=-\frac{\partial}{\partial z}\int_0^{R(z)}\!\! dr' r' W(z,r').
 \label{tada}
\end{equation}
Introducing the Stokes stream function $\Psi$, which we identify as the integral on the rhs of \eqref{tada}, 
with $rV=-\partial \Psi/\partial z$ and $rW=\partial \Psi/\partial r$, we
obtain
\begin{equation}
 \frac{\partial}{\partial t}\left(\frac{1}{2}R^2\right)=-\frac{\partial \Psi}{\partial z},
 \label{tada1}
\end{equation}
thus confirming the connection between the functions $F$ and $\Psi$.

\section{Hamiltonian Structure}
\label{Hamiltonian}

In this section we discuss some issues regarding a possible Hamiltonian formulation of the present system, for which
there is a vortex sheet strength with nontrivial dynamical evolution.  It is useful to contrast this case with that
of a system of discrete vortex rings and its continuous limit \cite{Meleshko}, for despite some fundamental distinctions involving 
conservations laws, there are common mathematical structures involved. 
As first shown by Dyson \cite{Dyson}, the Lagrangian dynamics of a discrete set of coaxial vortex rings with centers $Z_i$ on
the $z$-axis, with radii $R_i$ and circulation $\Gamma_i$, can be written as
\begin{subequations}
 \begin{align}
  R_i{\dot Z}_i &=\frac{\Gamma_i}{4\pi}\left(\log\frac{8R_i}{a_i}-\frac{1}{4}\right)+\sum_{j=1, j\neq i}^{N} 
  \frac{\Gamma_j}{2\pi}\frac{\partial I_{ij}}{\partial R_i}\label{ring1}\\
  R_i{\dot R}_i &=-\sum_{j=1, j\neq i}^{N} 
  \frac{\Gamma_j}{2\pi}\frac{\partial I_{ij}}{\partial Z_i}\label{ring2},
 \end{align}
\label{ring}
\end{subequations}
where $I_{ij}=\sqrt{R_i R_j}M(k_{ij})$ is the mutual inductance between the two loops $i$ and $j$. The first
term on the rhs of \eqref{ring1} is the approximate self-induced velocity of ring $i$, where $a_i$ is the core radius,
which serves as a cutoff for the localized induction approximation.

The evolution equation \eqref{fluxMaxwell} can now be seen as the continuum limit of \eqref{ring2}, expressed in an
Eulerian form.  A direct calculation shows a less obvious result, namely that the axial velocity $z_t$ in \eqref{CL2a} is 
the equivalent of \eqref{ring1},
without the self-induction term.  In fact,
\begin{equation}
r z_t = {1\over  \pi }{\partial \over \partial r}\int_{-\infty}^{\infty} d\alpha^{\prime}\gamma' {\sqrt {r r^{\prime}} M(k)}
={\partial F \over \partial r}.
\label{fluxMaxwell2}
\end{equation}
By analogy to the discrete case, in which
one can introduce an energy that is quadratic in the circulations,
$\Gamma_i$, namely
\begin{equation}
 U=\sum_{i=1}^{N}\sum_{j=1, j\neq i}^{N} 
  \frac{\Gamma_i\Gamma_j}{2\pi}I_{ij},
\end{equation}
and from which one obtains the equations of motion
\begin{subequations}
 \begin{align}
  \Gamma_i R_i{\dot Z}_i &=\frac{\Gamma_i^2}{4\pi}\left(\log\frac{8R_i}{a_i}-\frac{1}{4}\right)+\frac{\partial U}{\partial R_i} 
  \label{ring1a}\\
  \Gamma_iR_i{\dot R}_i &=-\frac{\partial U}{\partial Z_i}\label{ring2a},
 \end{align}
\label{ringa}
\end{subequations}
one can take the continuum limit to obtain \cite{Novikov}
\begin{equation}
 {\cal U}=\frac{1}{2\pi}\int\! d\alpha \int d\alpha' \gamma(\alpha)\gamma(\alpha')
  \sqrt{r r'}M(k),
\end{equation}
which, generalizing to functional derivatives, yields
\begin{equation}
\gamma r z_t =  {\delta {\cal U} \over \delta r} \ \ \ \ {\rm and} \ \ \ \ 
\gamma r r_t = -{\delta  {\cal U} \over \delta z}.
\label{pseudoHam}
\end{equation}
This is of the same form as the discrete dynamics \eqref{ringa}, but with the important distinction that the vortex
sheet strength $\gamma$ itself depends on time, whereas the individual circulations $\Gamma_i$ in the discrete case
do not.  Moreover, in the discrete case it is possible to rewrite the dynamics so the left-hand-sides are total time
derivatives, rendering them truly Hamiltonian.  In contrast, the dynamics \eqref{pseudoHam} do not obviously have
this feature.  The fact that \eqref{strength} can be throught of as a nonholonomic constraint may offer a path to
obtain an ``almost-Hamiltonian'' dynamics \cite{Fernandez}.

\section{Stability Analysis}
\label{stability}

In this section we show how the vortex sheet dynamics in the Eulerian form reproduces known stability results 
\cite{Alterman,Sterling} for
capillary jets, both with and without a streaming velocity within the fluid enclosed by the sheet.  Note that our
assumption that the fluids inside and outside the sheet have the same density precludes recovering the original
stability result of Rayleigh \cite{Rayleigh}, which assumed vacuum outside.

We first consider the case with a quiescent fluid on both sides of a vortex sheet of radius $R$, and linearize
the equations of motion for small perturbations in $\gamma$ and $r$ of the form 
\begin{equation}
\gamma = {\hat \gamma} e^{ iqz + \beta t} \ \ \ \ {\rm and} \ \ \ \  
r  = R +  {\hat \zeta} e^{ iqz + \beta t}.
\label{perts}
\end{equation}
At this order the mean curvature has the
simplified form $\kappa \simeq r_{zz}-r^{-1}$.  The resulting vortex sheet evolution equation is
\begin{equation}\label{strength_noflow}
 \beta {\hat \gamma} =  i P (1 - P^2){\sigma \over  \rho R^3}~{\hat \zeta},
\end{equation}
where $P=qR$.  Since, in the absence of background fluid motion, $\gamma$ is first order in the perturbation, the
remaining factors on the rhs of \eqref{fluxMaxwell} are those corresponding to a cylinder.  Let
\begin{equation}
k_R^2 = {4 R^2 \over  (z - z^{\prime})^2 + 4R^2}, \ \ \ \ {\rm and} \ \ \ \ x = {(z^{\prime}-z) \over 2R }.
\label{kRdefine}
\end{equation}
Then, the linearization of \eqref{fluxMaxwell} is
\begin{equation}
 \beta {\hat \zeta} =  -{4\over  \pi }i P {\hat \gamma} \int_{0}^{\infty}  M(k_R) \cos(2Px) ~dx
 \label{hatzeta}
\end{equation}
After calculating the integral (see Appendix), we substitute $\hat\zeta$ from \eqref{hatzeta} into
\eqref{strength_noflow} to obtain, in agreement with previous results \cite{Alterman,Sterling},  the growth rate
\begin{equation}
\beta^2 = {\sigma \over  \rho R^3}~ P^2\left(1 - P^2\right)I_1(P)  K_1(P)
\label{linstab1}
\end{equation}
where  $I_n$ and $K_n$ are the $n$-th order modified Bessel functions of the first and second kind, respectively.  

Demonstrating that the flux-form PDE reproduces the stability results in the presence of streaming flows, obtained by Alterman \cite{Alterman} and by Sterling and Sleicher \cite{Sterling}, follows the same procedure as the calculation above, 
but requires a more delicate analysis. 
This complexity is related to the limiting procedure of vanishing sheet thickness $\delta$ and
vanishing viscosity
used to arrive at the evolution equation for the vortex sheet strength \eqref{strength1}.  In particular, these limits 
preclude the determination of a a unique value of the tangential sheet velocity $z_t$.   For concreteness, consider the situation in which the streaming
velocities inside and outside of the cylindrical vortex sheet are $U$ and $0$, respectively.  In the 
Lagrangian formalism, this ambiguity is eliminated by setting  $z_t=U/2$ \cite{CaflischLi}.
However, this choice cannot satisfy 
the boundary conditions on either side of the sheet in the Eulerian frame. It is therefor necessary to reinstate the 
appropriate boundary 
conditions, which is equivalent to taking the limit $\delta \to 0$ at the end of the calculation. 

The calculation proceeds by systematic perturbation of the terms within the flux integral.  We expand theseparately
the two, obtaining 
\begin{equation}
M(k) = M(k_R ) + {dM \over dk}\bigg|_{k_R} (k - k_R) + \cdots \ \ \ \ {\rm and} \ \ \ \ 
\gamma^{\prime} {\sqrt {r r^{\prime}} } \simeq  R ( \gamma_0 + {\tilde \gamma}) \left( 1 + { \zeta^{\prime}+\zeta \over 2 R} \right) + \cdots,
\label{22}
\end{equation}
where $k_R$ is defined in \eqref{kRdefine}.
Using
\begin{equation}
k - k_R \simeq k_R (1 -k_R^2) \left( { \zeta^{\prime}+\zeta \over 2 R} \right),
\label{23}
\end{equation}
and collecting terms, we obtain (see Appendix)
\begin{equation}
\gamma^{\prime} {\sqrt {r r^{\prime}} } M(k) \simeq \gamma_0  R M(k_R) 
+ R {\tilde \gamma^{\prime}}M(k_R)-
  \gamma_0 R k_R [E(k_R) - K(k_R)] \left( { \zeta^{\prime}+\zeta \over 2 R} \right).
\end{equation}
Substituting into the equations of motion and retaining only first order terms, we find
\begin{equation}
\beta{\hat \zeta}=  - {1\over  \pi }\left[ 2 i P {\hat \gamma} \int_{0}^{\infty}  M(k_R) \cos(2Px) ~dx - 
i {P \gamma_0 
{\hat \zeta}\over R}
\int_{0}^{\infty} \left[ 1 + \cos(2Px)\right] k_R \left[E(k_R) - K(k_R)\right] ~dx \right]
\end{equation}
After some laborious calculation of these non-trivial integrals we obtain
\begin{equation}
\beta{\hat \zeta} =  - i P  I_1(P) K_1(P){\hat \gamma} - i {\gamma_0 
\over 2 R}  P\left[ 1 + P \left(I_0(P)K_1(P)  - I_1(P) K_0(P)\right) \right]~ {\hat \zeta}.
\end{equation}
Using the Bessel function identity $P(I_0 K_1 + I_1 K_0) = 1$, we 
further reduce this expression to
\begin{equation}\label{stability1}
\left( \beta + i {P \gamma_0 \over  R} K_1(P) I_0(P) \right) {\hat \zeta}=  -2 i P  I_1(P)K_1(P) ~ {\hat \gamma}
\end{equation}

This equation is the evolution for perturbations to the mean radius, i.e. the mean location of the vortex sheet: 
$ r = (r^+ + r^-)/2)$, where $r^+$ and $r^-$ are
the outer and inner radius of the infinitesimaly thin sheet, respectively ($r^+ - r^- = \delta \to 0$). 
To reinstate the proper boundary conditions we note that \eqref{stability1}
should be split into two parts, the contribution from $r^+$ and the one from $r^-$. To do so we rewrite \eqref{stability1} as
\begin{equation}\label{stabilityboth}
\left( P \left( \beta + i {P \gamma_0 \over  R}\right) K_1(P) I_0(P)+ P( \beta -0) K_0(P) I_1(P)\right){\hat \zeta} =  -{ i P \over 2}{\hat \gamma} K_1(P) I_1(P) 
-{ i P \over 2}{\hat \gamma} K_1(P) I_1(P) 
\end{equation}
In this equation all terms involving $I_0$ are connected to the inner region, that is $r^-$, while those involving $I_1$ correspond to the
outer one ($r^+$) yielding two equations, one for each region:
\begin{equation}\label{stability-}
P \left( \beta + i {P \gamma_0 \over  R}\right) K_1(P) I_0(P){\hat \zeta} =  -{ i P \over 2}{\hat \gamma} K_1(P) I_1(P)  
\end{equation}
and
\begin{equation}\label{stability+}
P  \beta {\hat \zeta} K_0(P) I_1(P) =  -{ i P \over 2}{\hat \gamma} K_1(P) I_1(P) 
\end{equation}
A similar expansion of the vortex sheet evolution equation \eqref{strength} yields for the inner region
\begin{equation}\label{strength-}
\left( \beta + i {P \gamma_0 \over  R}\right){\hat \gamma} =  {\sigma \over  \rho} { i P (1 - P^2)\over R^3}{\hat \zeta},  
\end{equation}
and for the outer one
\begin{equation}\label{strength+}
 \beta {\hat \gamma} =  {\sigma \over  \rho} { i P (1 - P^2)\over R^3}{\hat \zeta} .
\end{equation}

Substituting \eqref{strength-} and \eqref{strength+} into \eqref{stability-} and \eqref{stability+}, respectively, and adding the 
contributions to obtain the total expansion for $r = (r^-/2)+(r^+/2)$ yields
\begin{equation}\label{strength2}
\beta^2  + 2 i \beta \frac{\gamma_0}{R} P^2 I_0(P)K_1(P) =  
\frac{\sigma}{\rho R^3} P^2\left(1-P^2\right)I_1(P) K_1(P)+\frac{\gamma_0^2}{R^2}P^3 I_0(P)K_1(P)~,
\end{equation}
and which coincides with previous results \cite{Sterling,Alterman} for the case where the inner fluid velocity is  $U = \gamma_0 $ and 
the outer fluid is stationary.

\section{Derivation of a local PDE}
\label{sec:pde}
\begin{figure}[t]
\centering
\includegraphics[width=0.8\textwidth]{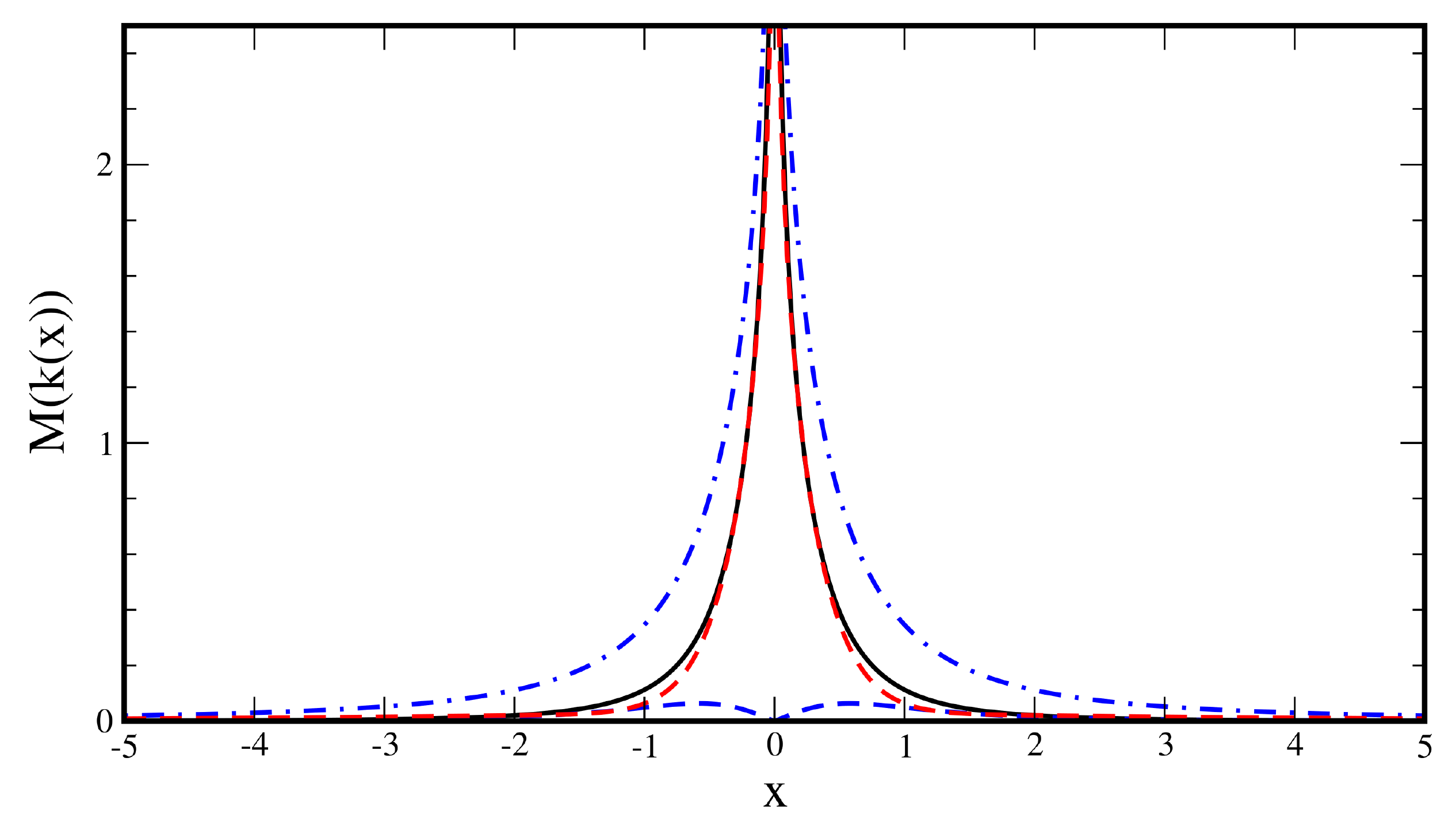}
\caption{Uniform approximation to the Maxwell function.  Solid black line is the exact expression
$M(k(x))$.  Dash-dot and dashed blue lines are the inner and outer approximations, respectively.
Dashred red line is the uniform approximation $M_u(k)$.}
\label{fig2}
\end{figure}

To obtain an approximate PDE that describes the dynamics of slender necks, 
it is necessary to find a suitable approximation to $M(k)$ that
would make it possible to calculate the integral in the flux equation in a controlled manner. With this purpose in mind,
it is very useful to 
note \cite{GradRyz} that $M(k) = Q_{1/2} (\chi)$, with $Q_{1/2} (\chi)$ the associated Legendre 
function of index $1/2$ with variable $\chi = (2/k^2) -1$,  
which obeys the differential equation 
\begin{equation} 
(1 -\chi^2) {d^2 Q \over d \chi^2} - 2 \chi {d^2 Q \over d \chi^2} + {3 \over 4}Q = 0.
\label{Legendre}
\end{equation} 
This identity makes it possible to obtain a uniform approximation for $M(k)$ by matching the inner
and outer solutions, ($\chi \to 1$ and $\chi \to \infty$, respectively).  In the inner region the limiting behavior is
\begin{equation}
Q_{1/2} \simeq A \ln \left(\sqrt{{2\over \chi-1}}\right) \ \ \ {\rm for } \ \ \ \chi \to 1, 
\end{equation}
while in the outer one is 
\begin{equation}
Q_{1/2} \simeq B \sqrt{ { (\chi+1)^3 (\chi-1) \over 16}}  \ \ \ {\rm for } \ \ \ \chi \to \infty ,
\end{equation}
where $A$ and $B$ are constants to be adjusted by matching the inner and outer solutions, with the final result in terms of $k$ given by
\begin{equation}
M(k) \simeq M_u(k)= {1\over 4} \left[ 3 \ln \left( {1\over (1-k^2)}\right)-
\pi k^{3/2} {\sqrt {1-k^2}}  \right].
\label{Muniform}
\end{equation}
The fit of $M(k)$ provided by the uniform approximation $M_u(k)$ is remarkable; $M(k)- M_u(k) \gtrapprox 0$ only in a 
small neighbourhood of $ k^2 = 1/2 $; everywhere else it
is almost impossible to distinguish $M_u(k)$ from the exact solution (see Fig. \ref{fig2}).

We can now approximate $M(k)$ by $M_u(k)$ in the evolution equation for $r^2(t)$ \eqref{fluxMaxwell}.
In order to expression $M_u(k)$ in a form that allows calculation of the relevant integral, we Further, 
using the expression for $k$ in terms of $z$, $z^{\prime}$, $r$, and $r^{\prime}$ it is possible to rewrite  in a way that
allows to calculate the integrals. In particular, using
\begin{equation}
k^2 = {4 r r^{\prime} \over {(z^{\prime}-z)^2 + (r + r^{\prime})^2}},
\end{equation}
the denominator inside the logarithm in \eqref{Muniform} is 
expandable in Taylor series in the limit $ z \to z^{\prime}$ because
the ratio $ (r^{\prime}- r)/(z^{\prime}-z) \to r_z$.  
In this limit, $ r \to r^{\prime}$, and rescaling $r = \epsilon R$ we find
\begin{subequations}
\begin{align}
\ln \left( {1\over (1-k^2)}\right) &\simeq \ln \left[ 
{(z^{\prime}-z)^2 + \epsilon^2(2R)^2 \over
 {(z^{\prime}-z)^2 \left(1 +  \epsilon^2 R_z^2\right)}} \right]  = \ln \left[ 
{x^2 + \epsilon^2\over
 x^2}\right] - \ln \left( 1 +  \epsilon^2 R_z^2\right), \\
k^2 {\sqrt{1-k^2}} &\simeq 8\epsilon^3 R^3
{|z^{\prime}-z|\left(1 +  \epsilon^2 R_z^2\right)^{1/2}\over {\left[(z^{\prime}-z)^2 + \epsilon^2 (2R)^2\right]^2}}
={\epsilon^3 |x|\over (x^2+\epsilon^2)^2} \left(1 +  \epsilon^2 R_z^2\right)^{1/2},
\end{align}
\end{subequations}
where $x$ is the same as defined in \eqref{kRdefine}.  The contribution to the flux $F$ arising from the
two terms in the uniform approximation \eqref{Muniform} involves $I_1$ and $I_2$, where, by the
symmetry of the integrand,
\begin{subequations}
\begin{align}
I_1 &= {3 \over 2} \epsilon \gamma R^2\int_0^{\infty}  
\ln \left[ {x^2 + \epsilon^2\over x^2}\right] dx + O(\epsilon^3 R R_z^2)\\
I_2 &=  \pi \epsilon^4 \gamma R^2 \int_0^{\infty} {x \over \left[ x^2 + \epsilon^2\right]^2} dx
+ O(\epsilon^3 R R_z^2)
\end{align}
\end{subequations}
Performing the integrations we obtain
\begin{equation}
I_1= {3 \pi \over 2} \epsilon^2 \gamma R^2 \ \ \ \ {\rm and} \ \ \ \ 
I_2 = {\pi \over 2} \epsilon^2 \gamma R^2 . 
\end{equation}

Collecting terms and substituting $r = \epsilon R$ in the LHS of the flux equation \eqref{fluxMaxwell} we obtain
\begin{equation}
{\partial R^2 \over \partial t}= -{\partial \over \partial z} \left(\gamma R^2 \right).
\label{lastone}
\end{equation}

While this result was obtained in a systematic way that would be useful in cases where it is necessary,
or desirable to obtain the next terms of the expansion, the procedure is not very intuitive. In fact, the 
meaning and procedence of \eqref{lastone} can be more easily understood by recognising that the first order approximation corrresponds to a cylindrical 
vortex sheet, where $r = r'$ and $\gamma' = \gamma$. Then, starting as before from \eqref{fluxMaxwell} we may write
\begin{equation}
{\partial r^2 \over \partial t} =  -{2\over  \pi } {\partial \over \partial z} \left( \gamma r^2 \int_{-\infty}^{+\infty} dx M(k(x))\right)~,
\label{cylinder}
\end{equation}
where we have used a variant of the usual variable $x$, namely $x= (z' -z)/2r$. Because the value of the integral  
is  $\pi/2$, \eqref{cylinder}
yields the same result \eqref{lastone} that was obtained for the first order term by employing the expansion method
making clear its geometric interpretation.

\section{Discussion} The present work has shown hwo the familiar axisymmetric vortex sheet dynamics can be recast exactly in
an Eulerian form which clearly displays the inherent flux form of the motion.  Previously established linear stability
results are easily recovered with this formulation.  Most signficantly, the Eulerian dynamics is well-suited to the development
of a systematic reduction to a local approximation when the aspect ratio of the system is suitably small.  This controlled 
approximation, which at leading order recovers well-known empirical results \cite{EggPont}, 
should lend itself to more rigorous studies both of the form and validity of the governing PDEs as well as
the singularities they produce.  As the underlying model includes the dynamics of an incompressible fluid surrounding
the vortex sheet, it consitutes a more physically relevant starting point for understanding singularity formation by
moving soap films than that provided by mean curvature flows.

In light of the large body of existing work on axisymmetric vortex sheets, there are many possible extensions of the present work,
including the introduction of swirl \cite{CaflischLiShelley} or helical symmetry \cite{CaflischLi}.  Greater challenges
would be to incorporate weak viscous effects \cite{Lundgren,Ceniceros}, and to complete the Hamiltonization of the inviscid Eulerian
case in the presence of surface tension.

\section{acknowledgments} We are indebted to Keith Moffatt for stimulating discussions 
at an early stage of this project, and to John Hinch for discussions on the asymptotic analysis and the significance
of the flux form.
This work was supported in part by Established Career Fellowship EP/M017982/1 from the EPSRC (REG \& AIP).
REG and AIP are grateful to the I.H.E.S., and especially Patrick Gourdon, for hospitality during an extended visit supported
by the Schlumberger Visiting Professorship (REG).  

\section{Appendix}

Here we show some intermediate steps in the stability analysis.   We begin from the first equation of
\eqref{22},
\begin{equation}
M(k) = M(k_R ) + {dM \over dk}\bigg|_{k_R} (k - k_R)
\end{equation}
where $k_R$ is as defined in \eqref{kRdefine}, and using the relationships between the elliptic functions
and its derivatives we find 
\begin{equation}
{dM \over dk} = {2 \over k^2} (E - K)  - {2\over k} \left[{dE \over dk} -{dK \over dk}\right] - K - k  {dK \over dk}.
\end{equation}
The corresponding first-order expansions of the relevant quantities are 
\begin{equation}
k^2 = {4 (R + \zeta^{\prime}) (R +\zeta ) \over  (z - z^{\prime})^2 + (2R +\zeta^{\prime}+ \zeta )) ^2 }
\simeq k_R^2 \left[ 1 + (1 -k_R^2) \left( { \zeta^{\prime}+\zeta \over R} \right)\right],
\end{equation}
which yields \eqref{23},
and 
\begin{equation}
{dM \over dk}\bigg|_{k_R} = {2 - k_R^2 \over k_R^2 (1-k_R^2)} E(k_R) - {2\over k_R^2}  K(k_R),
\end{equation}
where we have used the relationships $dE/dk = (E-K)/k$ and  $k dK/dk = E/(1-k^2) - K$ \cite{GradRyz},  and have substituted
 $k= k_R$. Introducing the convenient notation
\begin{equation}
M_0 \equiv M(k_R) \ \ {\rm and} \ \ M_1 \equiv  {dM \over dk}\bigg|_{k_R} k_R(1-k_R^2) ,
\end{equation}
we note that, after straightforward algebra, the quantity $ M_0 + M_1$, which will be needed later, reduces to $ -k_R (E-K)$.

Utilizing the second of the equations in \eqref{22} and collecting all the contributions we obtain
\begin{equation}
\gamma^{\prime} {\sqrt {r r^{\prime}} } M(k) \simeq \gamma_0  R M_0 + \gamma_0 R (M_0 + M_1) \left( { \zeta^{\prime}+\zeta \over 2 R} \right)
+ {\tilde \gamma} R M_0.
\label{62}
\end{equation}
Using again the identity $M(k_R) = Q_{1/2} (\chi)$, with argument $\chi = 2/k_R -1$, 
and substituting into \eqref{62} we obtain the first order result
\begin{equation}
\gamma^{\prime} {\sqrt {r r^{\prime}} } M(k) \simeq \gamma_0  R M(k_R) + R {\tilde \gamma^{\prime}} Q_{1/2}( {\chi}_R) -
  \gamma_0 R k_R [E(k_R) - K(k_R)] \left( { \zeta^{\prime}+\zeta \over 2 R} \right).
  \label{63}
\end{equation}
Substituting \eqref{63} into the equations of motion and retaining only first order terms we find
\begin{equation}
2 R {\partial \zeta \over \partial t} =  - {1\over  \pi } {\partial \over \partial z}\left[\int_{-\infty}^{\infty} d\alpha^{\prime}\gamma_0  M_0 R
 + R \int_{-\infty}^{\infty} d\alpha^{\prime} {\tilde \gamma} Q_{1/2} - \gamma_0
\int_{-\infty}^{\infty}  k_R [E(k_R) - K(k_R)]\left( { \zeta^{\prime}+\zeta \over 2 R} \right) ~d\alpha^{\prime} \right].
\label{perturbedr}
\end{equation}
The first integral on the rhs does not contribute to the equation of motion, for $\gamma_0=0$ 
identically when the 
fluids on both sides of the 
vortex sheet are stationary, and the integral is a constant when there 
is a streaming velocity in either one of the two regions.
The other integrals involve the Lagrangian parameter $\alpha$, and to be able to obtain the result in the Eulerian frame as a function of 
the coordinate $z$ we note that $ds = {\sqrt {r_{\alpha}^2 + z_{\alpha}^2}}~ d{\alpha}$ and $ds = {\sqrt {1 + r_z^2}}~ dz$.
Using these identities allows us to rewrite
\begin{equation}
dz = {\sqrt {1 + r_z^2 \over r_{\alpha}^2 + z_{\alpha}^2} }~ d{\alpha} \simeq (1 + {\cal O}(\zeta^2)) d{\alpha}
\end{equation}
making it possible to approximate $d{\alpha'} \simeq dz'$ to the required order. Replacing the expressions for the perturbations \eqref{perts} into\eqref{perturbedr}, rewriting
all exponentials in terms of $z' -z$ and expressing them in terms of sines and cosines makes it clear that the only non-zero contributions to
the integrals come from the cosine terms which are the only ones that produce even integrands. Then, 
defining $P = qR$, and performing the usual change of variables $x = (z'-z) /2R $ we find $k_R(x) = 1/(1+x^2)$, 
and $\chi(x) = 1+2 x^2 $, yielding the first one of the two stability equations
\begin{equation}
 \beta {\hat \zeta} =  - {1\over  \pi }\left[ 2 i P {\hat \gamma} \int_{0}^{\infty}  ~ Q_{1/2}(\chi(x)) \cos(2Px) ~dx - i {P \gamma_0 
{\hat \zeta}\over R}
\int_{0}^{\infty} \left[ 1 + \cos(2Px)\right] k_R [E(k_R) - K(k_R)] ~dx \right].
\label{horrible}
\end{equation}

The integrals have been calculated with the help of \cite{Okui} and \cite{GradRyz}. In particular, after a further change
of variable, $ y = {\sqrt 2} \ x$, the first integral acquires its canonical form 7.162.6 \cite{GradRyz} with $\nu = 1/2$, and $a = {\sqrt 2} \ P$, thus
\begin{equation}
\int_{0}^{\infty}  ~ Q_{1/2}(1+y^2) \cos({\sqrt 2} \  P y) ~dy = {\pi \over 2} K_1(P) I_1(P).
\end{equation}
The second integral can be divided in two contributions
\begin{subequations}
\begin{align}
 {\cal I}_1 &=  \int_{0}^{\infty} k_R [E(k_R) - K(k_R)] ~dx,  \\
 {\cal I}_2 &= \int_{0}^{\infty} \cos(2Px)~ k_R [E(k_R) - K(k_R)] ~dx.
 \end{align}
\end{subequations}

To calculate ${\cal I}_1$ we use the identity 3.2 (2) in \cite{Okui} for $p = 1$ and $a = x$ 
\begin{equation}
{ k_R \over 2 \pi p^2} [E(k_R) - K(k_R)]= \int_{0}^{\infty}  ~ e^{-2 p t}  N_0(a t) J_0(at) ~dt,
\end{equation}
where $N_0$ and $J_0$ are the Bessel functions. With this replacemente we obtain ${\cal I}_1 = - \pi /2$.
The second integral can be solved using identity 6.3(5) also in \cite{Okui}, for $a = x$ and $b =1$
\begin{equation}
{ k_R \over b^2} [E(k_R) - K(k_R)]= \int_{0}^{\infty}  ~ K_0(a t) J_1(at) ~dt , 
\end{equation}
to obtain
\begin{equation}
{\cal I}_2 = {\pi \over 2} \int_{0}^{\infty} {t J_1(t) \over {\sqrt{ 4P^2 + t^2}} } ~dt.
\end{equation}
This can be reduded to the standard form 6.565.1 with $a = 2P$  and $b = \alpha$ found in \cite{GradRyz} by noting that ${\cal I}_2$
can also be written as
\begin{equation}
{\cal I}_2 ={\pi \over 2} \lim_{\alpha \to 1} {\partial \over \partial {\alpha}} \left[ 
\int_{0}^{\infty} { J_0(\alpha t) \over {\sqrt{ 4P^2 + t^2}} } ~dt \right] = {\pi \over 2} P \bigg[ I_1(P) K_0(P) - I_0(P) K_1(P) \bigg]
\end{equation}

Finally, after collecting all the contributions and replacing them into \eqref{horrible} we find the first order equation
\begin{equation}
\beta {\hat \zeta}=  - i P {\hat \gamma} K_1(P) I_1(P) - i {P \gamma_0 
{\hat \zeta}\over 2 R}  \bigg[ 1 + P \left[ K_1(P) I_0(P) - K_0(P) I_1(P)\right] \bigg] .
\end{equation}
Making use of standard Bessel 
functions identities, it is possible to further reduce this expression to
\begin{equation}\label{stability1a}
\left( \beta + i {P \gamma_0 \over  R} K_1(P) I_0(P) \right){\hat \zeta} =  - i P  K_1(P) I_1(P) {\hat \gamma}.
\end{equation}

\bibliographystyle{aipauth4-1}

\end{document}